\documentclass[referee]{raa}

\usepackage{graphicx,times}
\usepackage{natbib}
\usepackage{amssymb,amsmath}
\bibpunct{(}{)}{;}{a}{}{,}

\usepackage[pagebackref=true]{hyperref}

\begin{document}

   \title{Quantized Redshift and its significance for recent observations
}

   \volnopage{Vol.0 (20xx) No.0, 000--000}      
   \setcounter{page}{1}          

   \author{Arindam Mal
      \inst{1}
   \and Sarbani Palit\
      \inst{2}
      \and Christopher C. Fulton\
      \inst{3}
   \and Sisir Roy
      \inst{4}
   }

   \institute{Indian Space Research Organization, Space Research Organization, Ahmedabad, India; {\it arindam.mal1982@gmail.com}\\
        \and
             Indian Statistical Institute, CVPR unit, Kolkata, India\\
             \and
             Protostar, Inc., 15652 Carrousel Drive, Canyon Country, CA, 91387, U.S.A\\
        \and
             National Institute of Advanced Studies, IISC Campus, Bangalore, India\\
\vs\no
}

\abstract{With the recent observational evidence in extra galactic astronomy, the interpretation of the nature of quasar redshift continues to be a research interest. Very high redshifts are being detected for extragalactic objects that are presumably very distant and young while also exhibiting properties that are characteristic of a more mature galaxy such as ours. According to Halton Arp and Geoffrey Burbidge, redshift disparities consist of an intrinsic component and are related to an evolutionary process. Karlsson observed redshift periodicity at integer multiples of $0.089$ in log scale and Burbidge observed redshift periodicity at integer multiples of $0.061$ in linear scale. Since Singular Value Decomposition (SVD) based periodicity estimation is known to be superior for noisy data sets, especially when the data contains multiple harmonics and overtones, mainly irregular in nature, we have chosen it to be our primary tool for analysis of the quasar-galaxy pair redshift data. We have observed a fundamental periodicity of $0.051$ with a confidence interval of $95\%$ in linear scale with the site-available Sloan Digital Sky Survey data release 7 (SDSS DR7) quasar-galaxy pair data set. We have independently generated quasar-galaxy pair data sets from both 2dF and SDSS and found fundamental periodicities of $0.077$ and $0.089$ in log scale with a confidence interval of $95\%$.
}

   \authorrunning{Mal.A. etal.}            
   \titlerunning{Quantized Redshift }  

   \maketitle

%
\newcounter{separate}
\setcounter{separate}{1}    

\ifnum \value{separate}=1
\clearpage
\else
\fi

\section{Introduction}           
\label{sect:intro}

The cosmological hypothesis defines the total observed redshift as:
\begin{equation}
(1+z)=(1+z_c)(1+z_{nc}), \label{eq:totalz}
\end{equation}

\noindent
where $z$ is the total observed redshift, $z_c$ is the redshift due to the cosmological contribution, and $z_{nc}$ is the redshift due to the non-cosmological contribution (\citealt{Bell1973}). According to standard theory, the redshifts of quasars are a result of spacetime expansion. If one plots quasar redshift against apparent brightness, one gets a non-linear relation, which implies that the linear Hubble relation is not valid for high redshift quasars (\citealt{Roy2000, Roy2007}).

\citealt{Hoyle1966} observed the physical association of quasars and galaxies by noting numerous instances of a quasar and a galaxy in close astrometric proximity but with different redshifts, for instance, a quasar with $z = 2.114$ very close to the nucleus of the galaxy NGC 7319 with $z = 0.022$. Arp observed many such pair cases in which filaments connect the objects, including prominent ones (e.g. NGC 4319 and MRK 205; NGC 3067 and 3C 232). Some quasars exhibit jets of unknown nature (e.g. 3C 345 in the vicinity of NGC 6212), while in some cases moving structures are radio sources accompanied by optical jets. \citealt{Lopez2004} observed two emission line objects with redshifts greater than $0.2$ in the optical filament apparently connecting the Seyfert galaxy NGC 7603 with $z = 0.029$ to its companion NGC 7603B with $z = 0.057$, a possible example of anomalous redshift. These findings lead to the conclusion that quasars are ejected from galactic nuclei and that quasar redshift is largely an intrinsic parameter. In the opening paragraphs of their introduction, \citealt{Fulton2012} directly compare and discuss the opposing viewpoints about cosmological versus quantized quasar redshifts. Using the first very large quasar and galaxy redshift surveys, 2dF quasars (\citealt{Croom2001}) and galaxies (\citealt{Sadler2002}), they support the quantized view with strong statistical evidence that high redshift quasars are indeed associated with low redshift galaxies. Standard theory predicts an evolutionary process in which super massive black holes (SMBH) begin their lives in the dust-shrouded cores of vigorously star-forming ``starburst'' galaxies before expelling the surrounding gas and dust and emerging as extremely luminous quasars. The recently observed GNz7q (\citealt{Fujimoto2022}) has exactly both aspects of the dusty starburst galaxy and the quasar. This observation lacks various features that are usually observed in very luminous quasars. The central black hole of GNz7q is still in a young and less massive phase at high redshift.

GN-z11 (\citealt{Jiang2021}) was photometrically selected as a luminous star-forming galaxy candidate at redshift $z > 10$ and is known to be the ``oldest galaxy''. The age of GN-z11 is estimated to be only $70$ million years and is moderately massive, suggesting that this young galaxy was born and grew rapidly, and the evidence of carbon and oxygen in GN-z11 indicates that this galaxy is not the first metal-free galaxy in the Universe and that it is a second generation galaxy. The detection of carbon and oxygen suggest special physical conditions not found in present day galaxies. Its accurate redshift remains unclear.

\ifnum \value{separate}=1
\clearpage
\else
\fi

Advancing a widely held conjecture that an SMBH must be the power source for any extremely massive extragalactic object, a search is on to find ultraluminous quasars to serve as examples. Part of this activity is to explain why the luminosity of each such object exceeds the Eddington limit (\citealt{Schindler2021}). Accompanying this scenario is the fact that if the quasars are actually at cosmological distances, they are by definition engaging in superluminal motion, as opposed to being simply luminous, moving at less than the speed of light, and much less distant. Observations of metallicity in high redshift quasars raise questions about the evolutionary process. Production of heavy elements by known nuclear processes occur at later stages of quasar evolution, but \citealt{Juarez2009} found high metal abundances in high redshift, presumably very young, quasars. They also found that the abundance of carbon relative to silicon and oxygen does not evolve significantly in quasars.

\citealt{arp1994} and \citealt{Arp2001} observed that a group of high redshift quasars appears to be physically connected with a lower redshift galaxy. They claimed that quasar redshifts do not indicate distance, that quasar redshifts are quantised, and that they obey a simple formula:
\begin{equation}
\frac{1 + z_{k+1}}{1 + z_k} = 1.227, \label{eq:karlsson}
\end{equation}

\noindent
where $z_k$ is the redshift of a quasar and $z_{k+1}$ is the next higher redshift. The factor $1.227$ was subsequently raised to $1.228$ in conjunction with using an anchor redshift of $1.960$ (\citealt{Fulton2012} Appendix). \citealt{Karlsson1977} observed that the peaks of the histogram of the redshifts form a mathematical series, $z = 0.061, 0.300, 0.600, 0.960, ...$. \citealt{Hawkins2002} found non-existence of quasar periodicity, but their methodology was challenged by \citealt{Napier2003}. \citealt{Tang2005, Tang2008} claimed non-existence of periodicity with a data set that was fifteen times larger than the previous one, but their work was directly contradicted by \citealt{Fulton2012}, and subsequently by \citealt{Fulton2018}, demonstrating redshift periodicity with very high statistical significance using ejection velocity
computations. \citealt{Mal2020} and \citealt{Mal2022} have already shown the existence of redshift periodicity for quasars as well as for galaxies. Here we have analysed quasar, galaxy, and quasar-galaxy pair data sets in an effort to ultimately understand the physics behind redshift quantisation.


\section{Methodology}
\label{sec:methd}

\subsection{Proposed method}
\label{subsec:propmethd}

We examine the existence of redshift periodicity in variations of five redshift data sets following the procedure outlined here. After the initial selection of data in accordance with any flag values provided, selection of the bin width for formation of the histogram is optimised as in \citealt{Shimazaki2007, Shimazaki2009}, minimising the overall cost function for each overall data set. All the data sets are originally sourced by other related work from the Sloan Digital Sky Survey data release 7 (SDSS DR7) or from the 2dF.

\ifnum \value{separate}=1
\clearpage
\else
\fi

An SVD-based method (\citealt{Kanjilal1995}) has been adapted for estimating the fundamental periodicity present in the histogram of the redshift data. \citealt{Mal2020} established the superiority of the SVD-based approach of periodicity detection over the periodogram-based approach. The periodogram is not a suitable tool for data of a quasi-periodic nature or a data set containing multiple periodic components and a large number of overtones or a somewhat irregular periodic part. Hence, periodicity detection for the data sets examined in this paper has been performed using the SVD-based method.

\subsection{Description of data sets}
\label{subsec:desc}

We have analysed data sets from different sources. First, the SDSS DR7 database contains information pertaining to quasar-galaxy pairs, where each quasar is projected within 100 kpc of a galaxy and has a redshift larger than the galaxy redshift by an amount equal to the total error in the redshift obtained from the quasar spectrum. The quasars and galaxies with misidentified Lyman-$\alpha$ forest lines are omitted. A total of $97,489$ quasar-galaxy pairs are reported from a sample of $10,5783$ spectroscopic quasars and $798,948$ spectroscopic galaxies. This data set contains spectroscopically observed quasar-galaxy projections that can be used to study quasar absorption-line systems arising from known galaxies with $0 < z < 0.6$ and quasars with $0 < z < 3.5$. The quasar and galaxy redshift data was collected from SDSS DR7 and sifted to find galaxies with definite Ca-II and Na-I absorbers by \citealt{Cherinka2011}.

Second, \citealt{Fulton2012} examined the 2dF quasar-galaxy data set. Statistical data sets were prepared separately for galaxies and quasars as described in Section 2 therein. Examinations were performed in terms of the ejection hypothesis using a procedure that, though wider in scope, is equivalent to examining quasar-galaxy pairs. The results not only confirmed the ejection hypothesis and quasar periodicity but also the existence of many quasar-galaxy pairs with disjoint redshifts and at high significance for both conjectures. The data set contains $7,849$ quasar-galaxy pairs for which each quasar is within $30'$ of its one and only paired galaxy.

Third, \citealt{Fulton2018} examined the SDSS DR7 quasar-galaxy data set in a manner procedurally identical to that just described for the 2dF data set.

Fourth, the same \citealt{Fulton2018} SDSS DR7 data set can also be examined as a variation in which the quasar-galaxy pairs are subjected to ejection model periodicity constraints, thus culling the data set to contain only pairs for which the quasar is found to be physically associated with only one galaxy.

Fifth, the SDSS AGN data set examined by \citealt{Liu2019} can be treated as the galaxy portion of a quasar-galaxy data set that is then paired with the SDSS DR7 quasars from \citealt{Fulton2018}.

\subsection{Data extraction}
\label{subsec:datext}

\citealt{Narlikar1993} explain that if quasars are physically connected with a parent galaxy, the redshift of each quasar must be transformed to the reference frame of the putative parent galaxy (\citealt{Narlikar1980}):

\ifnum \value{separate}=1
\clearpage
\else
\fi

\begin{equation}
1 + z_0 = \frac{1 + z_q} {1 + z_g} \label{eq:putative}
\end{equation}

\noindent
The computations used to examine Karlsson redshift periodicity in quasars and to implement the ejection hypothesis are detailed in the appendix of \citealt{Fulton2012}, and a complete description of the quasar family detection procedure is given in Section 5 of \citealt{Fulton2018}. The algorithm that performs the periodicity analyses for both papers is also capable of generating quasar-galaxy pair data sets. Using quasars alone, galaxies alone, or adding pairs to the mix allows the examination of individual data sets with three different redshift values:

\begin{itemize}

\item{\textbf{$z_q$:} This is the measured redshift obtained by a survey for a quasar.}

\item{\textbf{$z_g$:} This is the measured redshift obtained by a survey for a galaxy.}

\item{\textbf{$z_0$:} This is the transformed redshift of a quasar. The measured redshift obtained by a survey for the quasar is transformed to the rest frame of a paired galaxy using the measured redshift of the galaxy. Only the transformed redshift of the quasar is under consideration. The measured redshifts of the quasar-galaxy pair are unknown to the analysis and are used only to compute the quasar $z_0$ when creating the redshift data set.}

\end{itemize}

The data extraction varies somewhat with the data set as indicated in Section \ref{subsec:desc}, and there are additional differences as follows in the redshifts being examined:

\begin{itemize}

\item{\textbf{Quasars:} Only quasar redshifts are being considered, and as described in the redshift value list above, the redshift can be either $z_q$ or $z_0$. If the redshifts are $z_0$ values, the data set follows the prescription below for quasar-galaxy pairs.}

\item{\textbf{Galaxies:} Only galaxy redshifts are being considered, and as described in the redshift value list above, the redshift can be only $z_g$.}

\item{\textbf{Quasar-galaxy Pairs:} Each pair under consideration is one quasar and one galaxy, and as described in the redshift value list above, the redshift can be only $z_0$. No quasar-quasar or galaxy-galaxy pairs are being considered. The list of pairs for a given data set is determined by an astrometric association algorithm that for each galaxy determines which quasars lie within $30'$ of that galaxy. Each quasar so positioned is considered a pair with the galaxy independently of each of the other quasars that are also paired with that same galaxy. The algorithm determines whether each quasar is paired with more than one galaxy or is uniquely associated with only one, so that the list of pairs can respectively be generated either with all pairs or with unique pairs only.}

\end{itemize}

\subsection{The procedure for analysis}
\label{subsec:procedure}

As proposed by \citealt{Mal2020}, the two main stages of the approach consist of the formation of an appropriate histogram after determining the optimal bin width and application of SVD for periodicity determination.

\ifnum \value{separate}=1
\clearpage
\else
\fi

\noindent
For convenience, the procedure is briefly outlined here.

\noindent
\textbf{Determination of optimal bin width and histogram formation:} The optimal bin width is obtained by minimising the mean integrated square error (\citealt{Shimazaki2007, Shimazaki2009}) for the entire data set, and that bin width is used for the analysis of the redshift data.

Bin width has an impact on histogram data analysis, so proper bin width estimation is important. For this paper we have made use of simulation with random numbers and periodic numbers and observed that there is no periodicity in the random number case but in the other case a proper period length is found. The optimum bin width $\Delta$ of the histogram is obtained using the following formula:

\begin{equation}
C(\Delta) = 2K - v / {\Delta}^2, \label{eq:delta}
\end{equation}

\noindent\\
where $C$ is the cost function, $K$ is the mean of the data, and $v$ is the variance of the data. The optimal bin size is determined using minimisation of the cost function. The input data varies with random sampling for 1000 iterations, and the bin size and resulting sensitivity vary accordingly, which is defined as the confidence interval. The mean integrated square error is computed as:

\begin{equation}
MISE = 1 / T \int_{0}^{T} E({\lambda}’ - {\lambda})^2 dt \label{eq:mise1}
\end{equation}

\noindent\\
where $MISE$ is defined by how close the estimator ${\lambda}’$ is to observations over the time period $T$, and $E$ is the expectation parameter.

\noindent\\
\citealt{Shimazaki2007} adapted the above equation for optimisation of the histogram band width by segmenting the total observation period $T$ into $N$ intervals of size $\Delta$, with the $MISE$ redefined as:

\begin{equation}
MISE = 1 / \Delta \int_{0}^{T} 1 / N  \sum_{i=1}^{N} \{E(\theta - {\lambda}t)^2\} \label{eq:mise2}
\end{equation}

\noindent
This equation is solved for $\Delta$ to obtain the cost function using equation \ref{eq:delta}.

The existence of periodicity in a data sequence is examined using the Singular Value Decomposition (SVD) and a matrix formulation.

\noindent
\textbf{Matrix formulation \& application of SVD}: In order to examine the existence of a periodic component in the data sequence, candidate period lengths are selected and the input matrix for application of SVD is formed as follows. For a candidate period length of say, $L$, the corresponding data matrix $A$ is formed by partitioning the data into contiguous segments of each length $L$ and placing each segment (aligned in phase) as a row of $A$ such that $A$ is of size $M \times L$. Singular value decomposition of $A$ is computed to yield the singular values $\sigma_1, \sigma_2,\cdots,\sigma_p$ arranged in decreasing order of magnitude, where $p$=min(M,L) and is also the rank of $A$.

\ifnum \value{separate}=1
\clearpage
\else
\fi

It may be noted that if the data sequence under consideration is indeed perfectly periodic with period length $L$, then $A$ will have rank one and hence only one singular value {\it i.e.} $\sigma_1 \ne 0$ while $\sigma_i=0$ for $2\le i \le p$. If the data sequence contains a large periodic component of length $L$ and smaller noise-like signals, the rank of $A$ would no longer be one as all the $p$ singular values would be non-zero. The magnitude of $\sigma_1$ would still be much greater than $\sigma_2$ and the rest of the singular values. Hence, the distribution of magnitudes of the singular values of $A$ can be conveniently used to assess the degree of periodicity since in reality, considering the redshift sequences under analysis, not only may there be a multitude of different components (including possibly a periodic component), there is also noise arising from measurement errors or even on account of the data travelling over immense lengths of space and time. Similar to \citealt{Kanjilal1995}, we follow two approaches using two different indicators or measures for periodicity detection, as outlined below:

\begin{enumerate}
\item \textbf{Our first method}, SVR1, is based on the following measure of periodicity:
$$SVR1 = ({\sigma_1 \over \sigma_2})^2$$
A plot of this ratio versus row length is called the SVR1 plot which shows the presence of multiple peaks at integer multiples of the value of the period length of the periodic component contained which, for the case under consideration, is the fundamental redshift. It may be observed that for a perfectly periodic signal $SVR1 \rightarrow \infty$ while for a signal with no periodicity implying that $A$ is full rank, $SVR1=1$. Hence $1\le SVR1 < \infty$.

\noindent
\item \textbf{Our second method}, SVR2 (\citealt{Kanjilal1995}), is based on a quantification of the aperiodicity of the data sequence being analyzed :
\begin{equation}
\delta_{aper} = {(\sum_{i=2}^p\tau_i)^2\over 1+\sum_{i=2}^p \tau_i^2} \quad \mbox{where} \quad \tau_i = {\sigma_i \over \sigma_1}
\end{equation}
Then, SVR2 is defined as:
\begin{equation}
\mbox{SVR2}= \left({1\over \delta_{aper}}\right) {(p-1)^2\over p}
\end{equation}
The construction of SVR2 is such that it has the same range of values as SVR1, \textit{i.e.} $[1,\infty]$. It may be noted that it reflects the ratio of residual energy to the energy content of the aperiodic signal. The plot of this quantity versus row length is called the SVR2 plot which also shows repeated peaks at integral multiples of the value of the period length of the periodic component contained, if any.
\end{enumerate}
In each case, the presence of a periodic component is confirmed by the presence of a peak at the relevant period length as well as its integral multiples. In practice, there may be very slight deviations due to the presence of noise, as already explained above. For example, the peak at a particular integral multiple may be negligible or may occur at a shift of one or two positions from the integral multiple. Generic simulations are provided in Appendix A.

\clearpage

\section{RESULTS OF REDSHIFT PERIODICITY}
\label{sec:res}

\begin{table*}
\begin{center}

\caption{Results of redshift data sets.}

\label{tab:table2}

\begin{tabular}{rccccrrcrr} \\
\hline\hline
Data        & Extraction     & Object & Redshift            & Bin                  & Method & Fundamental & Confidence          & Confidence \\
 Set        &                & Type   & Range               & Width                &        & Period   & Interval               & Level      \\
\hline
   1$^{a}$  & SDSS$^{g}$     & Pair   & $0.08 < z_0 < 2.20$ &             $0.0044$ & SVR1   & $0.0510$ & $[0.0470$ to $0.0980]$ &   $95\%$   \\
            &                &        & $                 $ &                      & SVR2   & $0.0970$ & $[0.0930$ to $0.0980]$ &   $95\%$   \\
            & SDSS$^{e}$     & Quasar & $0.08 < z_q < 2.20$ &             $0.0042$ & SVR1   & $0.0640$ & $[0.0550$ to $0.0970]$ &   $95\%$   \\
            &                &        & $                 $ &                      & SVR2   & $0.0930$ & $[0.0920$ to $0.0970]$ &   $95\%$   \\
            & SDSS$^{f}$     & Galaxy & $0.03 < z_g < 0.10$ &             $0.0001$ & SVR1   & $0.0040$ & $[0.0020$ to $0.0090]$ &   $95\%$   \\
            &                &        &                     &                      & SVR2   & $0.0040$ & $[0.0030$ to $0.0100]$ &   $95\%$   \\
\hline
   2$^{b}$  & 2dF$^{h}$      & Pair   & $     log(1 + z_0)$ & $0.0020$ to $0.0070$ & SVR1   & $0.0770$ & $[0.0260$ to $0.0790]$ &   $95\%$   \\
            &                &        & $                 $ &                      & SVR2   & $0.0462$ & $[0.0260$ to $0.0790]$ &   $95\%$   \\
            & 2dF$^{f}$      & Galaxy & $       z_g > 0.03$ & $0.0010$ to $0.0010$ & SVR1   & $0.0150$ & $[0.0150$ to $0.0150]$ &   $95\%$   \\
            &                &        & $                 $ &                      & SVR2   & $0.0430$ & $[0.0420$ to $0.0690]$ &   $95\%$   \\
\hline
   3$^{c}$  & SDSS$^{h}$     & Pair   & $     log(1 + z_0)$ & $0.0270$ to $0.0530$ & SVR1   & $0.0884$ & $[0.0180$ to $0.0900]$ &   $95\%$   \\
            &                &        & $                 $ &                      & SVR2   & $0.0884$ & $[0.0230$ to $0.0910]$ &   $95\%$   \\
            & SDSS$^{f}$     & Galaxy & $       z_g > 0.03$ &             $0.0011$ & SVR1   & $0.0159$ & $[0.0117$ to $0.0244]$ &   $95\%$   \\
            &                &        & $                 $ &                      & SVR2   & $0.0797$ & $[0.0234$ to $0.0829]$ &   $95\%$   \\
\hline
   4$^{c}$  & SDSS$^{hi}$    & Pair   & $     log(1 + z_0)$ & $0.0010$ to $0.0030$ & SVR1   & $0.0877$ & $[0.0420$ to $0.0910]$ &   $95\%$   \\
            &                &        & $                 $ &                      & SVR2   & $0.0877$ & $[0.0430$ to $0.0900]$ &   $95\%$   \\
\hline
   5$^{cd}$ & SDSS AGN$^{h}$ & Galaxy & $              z_g$ &            $0.00006$ & SVR1   & $0.0220$ & $[0.0068$ to $0.0248]$ &   $95\%$   \\
            &                &        &                     &                      & SVR2   & $0.0140$ & $[0.0136$ to $0.0143]$ &   $95\%$   \\
\hline

\end{tabular}
\end{center}

$^{a}$Sourced from \citealt{Cherinka2011}.\\
$^{b}$Sourced from \citealt{Fulton2012}.\\
$^{c}$Sourced from \citealt{Fulton2018}.\\
$^{d}$Sourced from \citealt{Liu2019}.\\
$^{e}$Quasar redshift data set.\\
$^{f}$Galaxy redshift data set.\\
$^{g}$Quasar-galaxy pair redshift data set after quasar redshift transformation to a galaxy rest frame.\\
$^{h}$Quasar-galaxy pair redshift data set after quasar redshift transformation to a galaxy rest frame and $log(1 + z_0)$ taken.\\
$^{i}$Physically associated quasar-galaxy pairs.\\
\end{table*}

Table \ref{tab:table2} summarises the results of quasar, galaxy, and quasar-galaxy pair analyses. In each case, the fundamental period of the periodic component present is determined as the product of the peak location of the histogram and the optimal bin width. A peak location is selected as the base period if peaks are also observed at integral multiples of that peak location. The $95\%$ confidence level of the base period is determined using resampling of the redshift data set. A Monte Carlo type simulation is conducted over $1000$ iterations to compute the confidence interval.

\ifnum \value{separate}=1
\clearpage
\else
\fi

The redshift periodicity of a pair data set is computed using the $z_0$. The redshift periodicity of a quasar or galaxy data set is computed using the measured redshift, $z_q$ or $z_g$ respectively, instead of $z_0$, i.e. without transformation to a galaxy rest frame. The SVR spectrum exhibits the periodic peaks shown in the respective figures. In all figures, the primary series peaks are marked with a blue vertical line and a secondary strong peak, if present, is marked with a red vertical line. Depending on the data set, a periodic component, or in some cases two components, can be observed for different ranges of redshifts of quasars, galaxies, or quasar-galaxy pairs. Working by example, some salient observations are made about a specific result for each data set. Table \ref{tab:table2} summarises the results of quasar, galaxy, and quasar-galaxy pair analyses.

\begin{figure*}
\centering
\includegraphics[scale=0.6,angle=0]{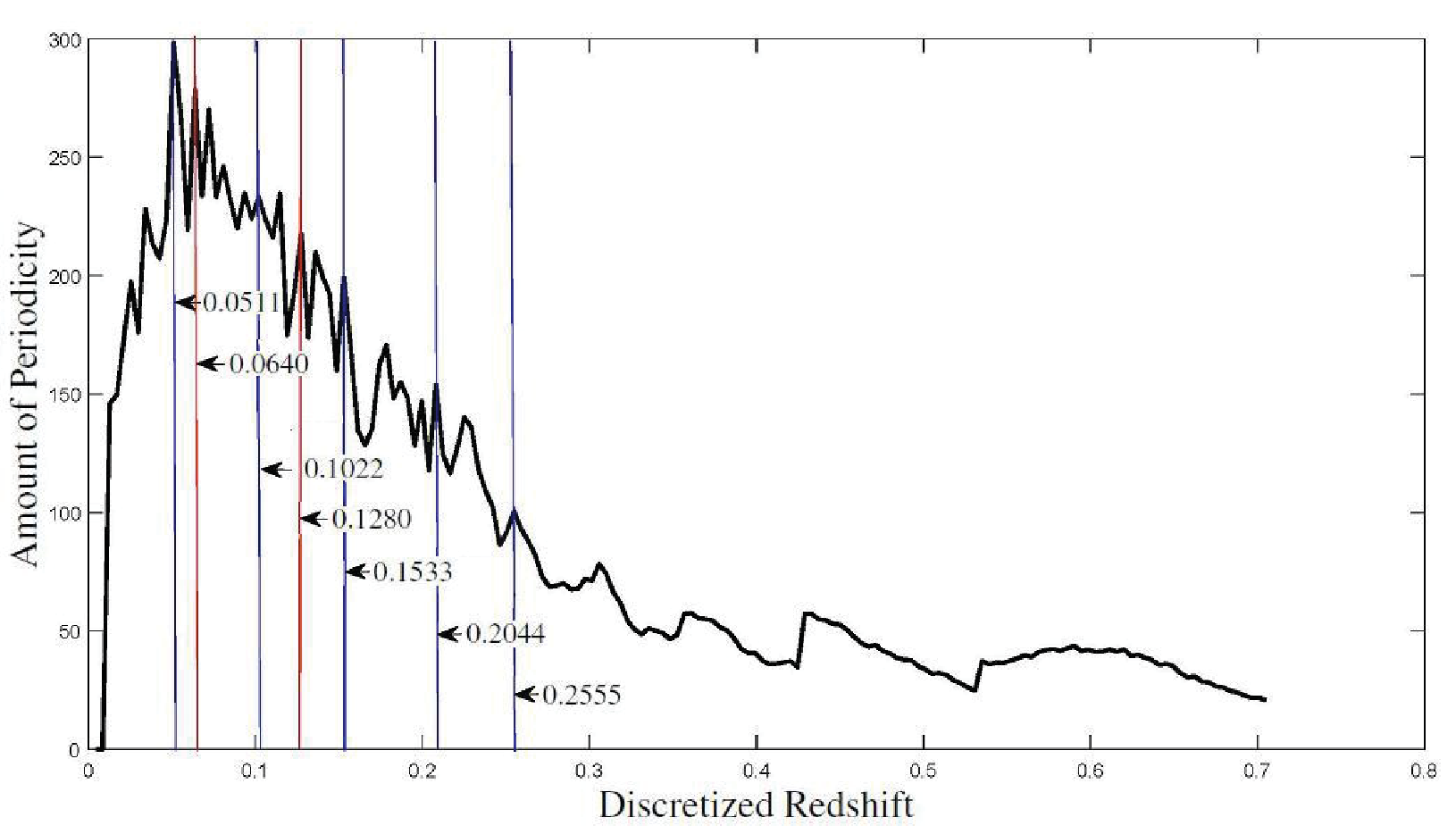}
\caption{SVR1 spectrum for quasars with $0.08 < z_0 < 2.20$.}
\label{fig:figure11}
\end{figure*}

The first data set is an extraction of \citealt{Cherinka2011} SDSS quasar-galaxy pairs. For the quasars with $0.08 < z_0 < 2.20$, Fig. \ref{fig:figure11} is an SVR1 spectrum showing a primary strong peak at $0.0510$, a secondary strong peak at $0.0640$, and additional peaks occur at multiples of both the primary and secondary peaks.

\begin{figure*}
\centering
\includegraphics[scale=0.6,angle=0]{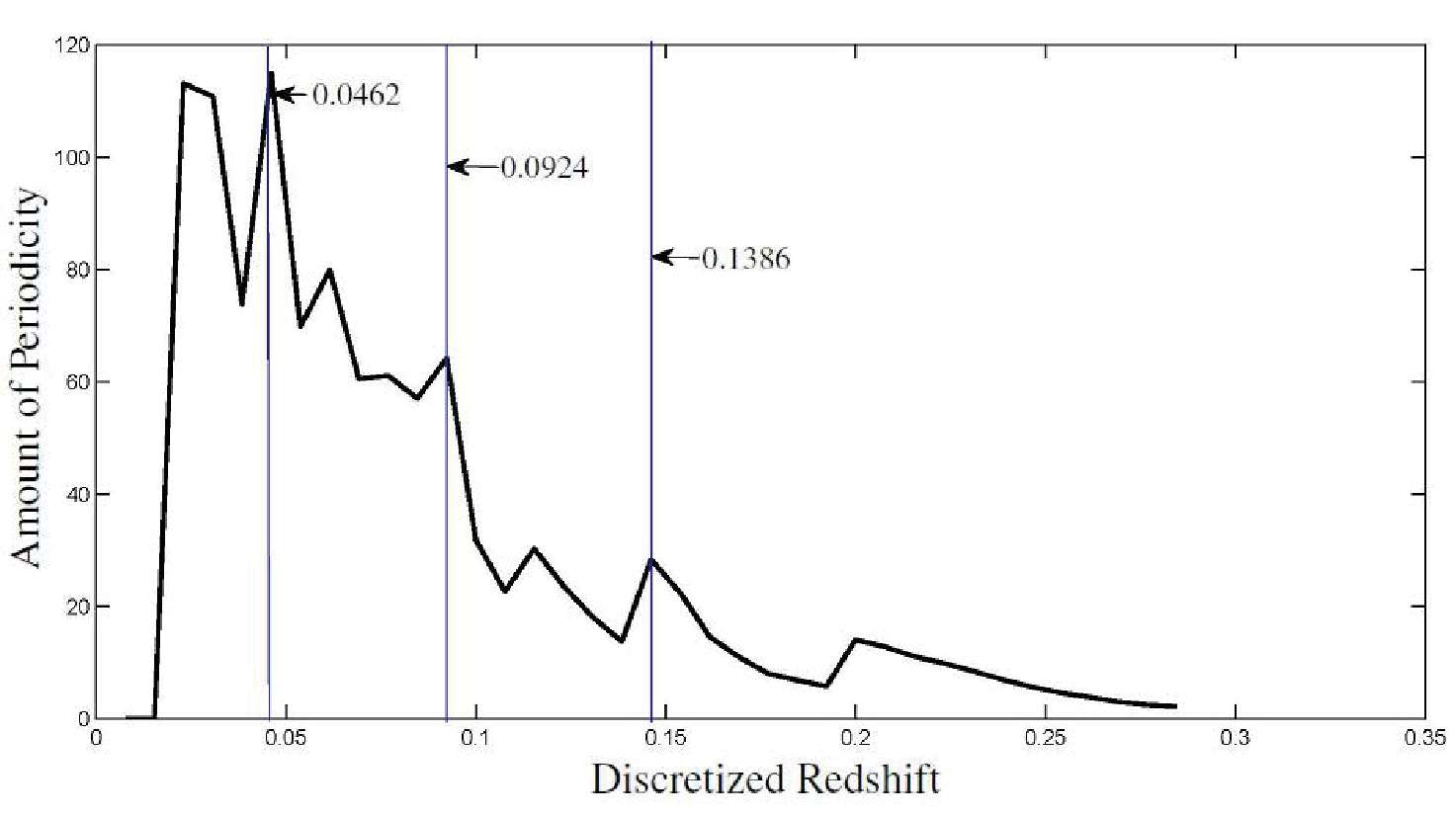}
\caption{SVR2 spectrum for 2dF quasar-galaxy pair $z_0$ values.}
\label{fig:figure12}
\end{figure*}

The second data set is an extraction of 2dF quasar-galaxy pairs from \citealt{Fulton2012} Section 2 and \citealt{Fulton2018} Section 5. For quasar-galaxy pair $z_0$ values, Fig. \ref{fig:figure12} is an SVR2 spectrum showing a primary strong peak in multiples of $0.0462$.

\begin{figure*}
\centering
\includegraphics[scale=0.6,angle=0]{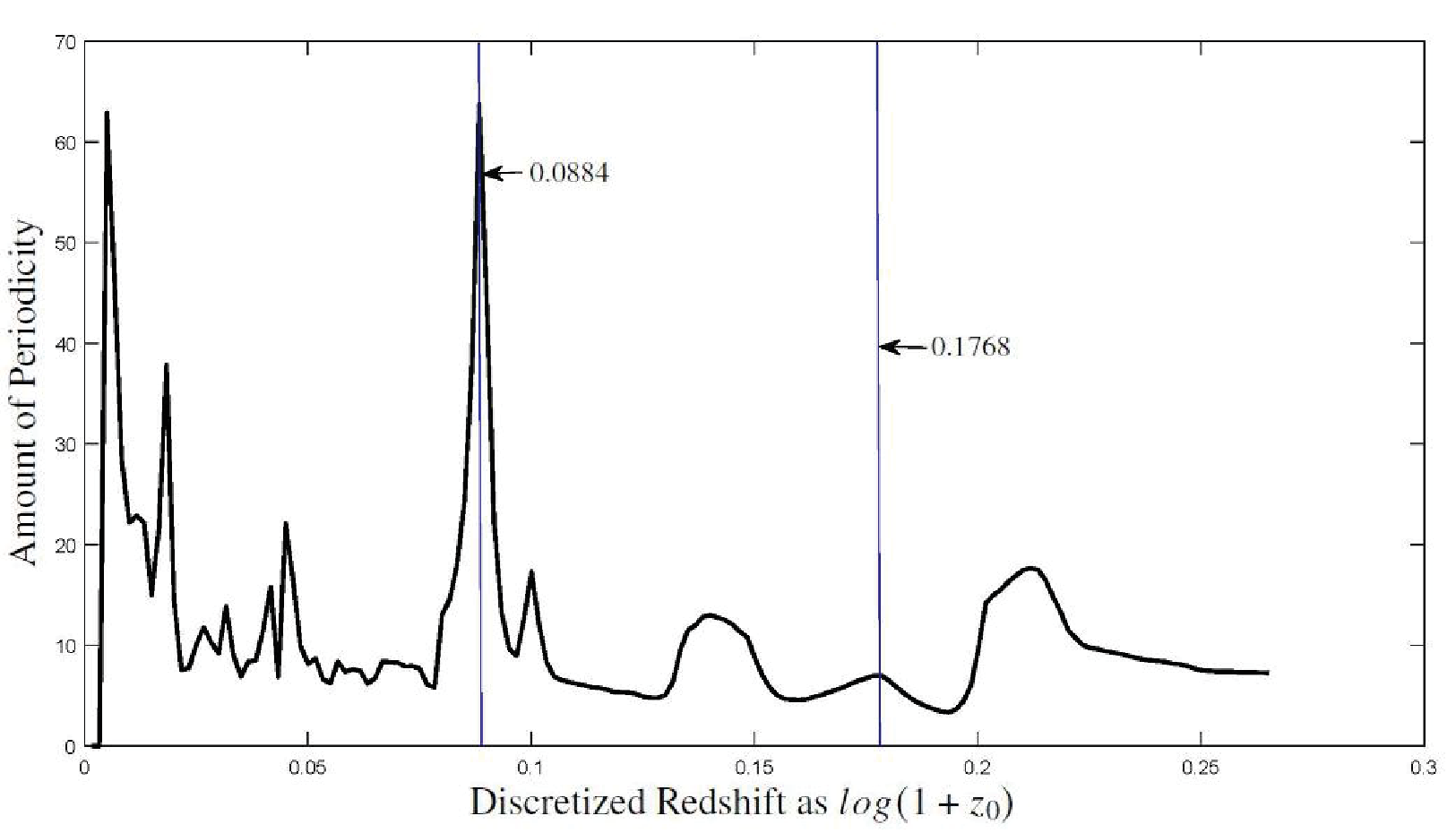}
\caption{SVR2 spectrum for SDSS quasar-galaxy pair $log(1 + z_0)$ values.}
\label{fig:figure13}
\end{figure*}

The third data set is an extraction of  SDSS quasar-galaxy pairs from \citealt{Fulton2018} Sections 2 and 5. For quasar-galaxy pair $log(1 + z_0)$ values, Fig. \ref{fig:figure13} is an SVR2 spectrum showing a primary strong peak in multiples of $0.0884$.

\begin{figure*}
\centering
\includegraphics[scale=0.6,angle=0]{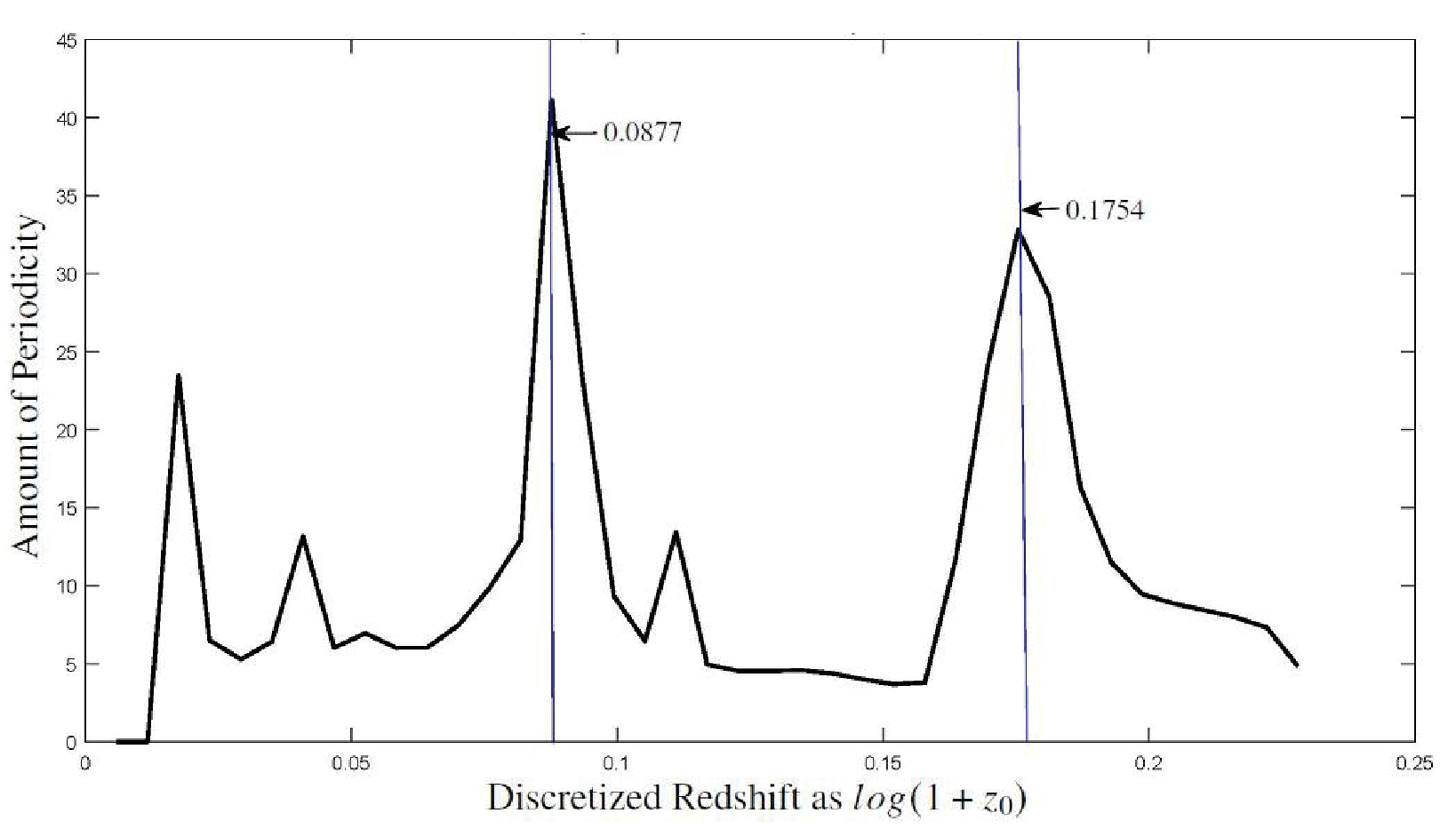}
\caption{SVR2 spectrum for SDSS-AGN and SDSS quasar quasar-galaxy pair $log(1 + z_0)$ values.}
\label{fig:figure14}
\end{figure*}

The fourth data set is an extraction of  SDSS quasar-galaxy pairs from \citealt{Fulton2018} Sections 2 and 5. This extraction is the same as in the third data set example except that it shas been modified to include only those pairs for which the generating algorithm has determined that the quasar and galaxy of each pair are physically associated.For quasar-galaxy pair $log(1 + z_0)$ values, Fig. \ref{fig:figure14} is an SVR2 spectrum showing a primary strong peak in multiples of $0.0877$.

The fifrh data set is an extraction of  SDSS AGNs (\citealt{Liu2019}) and  SDSS quasars from \citealt{Fulton2018} Sections 2 and 5 to form quasar-galaxy pairs. The fundamental periodicity repeat with $0.022$ using SVR1 method.

\ifnum \value{separate}=1
\clearpage
\else
\fi

It has been observed that if the noise level is increased by 4 dB at an arbitrary point in a periodic quasar/galaxy redshift data set, then, at a point later in the data set, the periodicity will disappear, amounting to the null condition.

\ifnum \value{separate}=1
\clearpage
\else
\fi

The redshift periodicity of quasars, galaxies, quasar-galaxy pairs may indicate that evolution of quasars into galaxies has occurred over time, and that is consistent with statistical tests of the ejection hypothesis (\citealt{Fulton2012}; \citealt{Fulton2018}).

\ifnum \value{separate}=1
\clearpage
\else
\fi

\section{Discussion}
Statistical analysis of redshift data clearly establishes the existence of discrete redshifts using quasar-galaxy pair data sets (\citealt{Fulton2012, Fulton2018}). We are therefore looking for models that are capable of explaining the periodicity of extragalactic redshifts. One such model is proposed by Hoyle-Narlikar (\citealt{Narlikar1980, Hoyle1964, Hoyle1966}) which considers periodicity of redshifts based on the Variable Mass Hypothesis (VMH). According to this model, the inertia of matter arises due to its interaction with other matter and the inertial matter is created from a zero mass surface based on a quantum principle. The excess redshift does not arise from high speed of ejection but from the low mass of the newly created matter. \citealt{Narlikar1980} explain that an ejected quasar can be bound to the parent galaxy with typical separations on the order of $100-200$ kpc. \citealt{Burbidge2000} explain redshift periodicity using VMH. Redshift periodicity is a direct confirmation of the fact that quantisation in redshift implies quantisation in mass. VMH theory predicts that the age of an object is usually measured from the zero mass epoch on its world line, so the higher the redshift of a quasar the younger it is. This is well matched with recent observations of the extragalactic objects HD1 and HD2 (\citealt{Harikane2022}), GNz7Q, and GNz11, and it should be noted that HD1 and HD2 are not likely to be lensed objects (\citealt{ZheLee2022}).

Another model proposed and elaborated upon by \citealt{Roy2000, Roy2007} based on the Wolf mechanism is known as Dynamic Multiple Scattering (DMS; see \ref{sec:dms}) theory. \citealt{Roy2000} have shown that DMS can explain the redshift for quasar-galaxy association, which depends on the properties of the environments around the galaxies as well as quasars. It raises lot of interest within the community about the alternate mechanism of redshift even in the absence of relative motion between the object and the observer. No works have so far been done regarding the existence of periodicity of redshift within this framework. However, we find that the redshift in this framework depends on the parameters directly related to the characteristics of the medium, i.e. correlation coefficents and the characteristic lengths associated with random refractive index of the environments around galaxies and quasars may explain the redshift periodicity. We emphasize that this can be verified in laboratory experiments as well.

\citealt{Lehto1990} developed a theoretical model which could predict redshift periodicity, in the framework of matter using three-dimensional quantised time. The time unit is Planck time. The redshift quantisation can be obtained based on the assumption that distances are quantised in Planck’s units. However,It is difficult to understand the concept of three-dimensional quantised time from the standard interpretation of modern physics.

\citealt{Wolf1986} explains correlation induced spectral shift as well as the broadening and shifts of spectral lines. \citealt{Roy2000, Roy2007} analyzed statistically the V\'eron-Cetty (V-C) quasar catalogue (\citealt{Veron2006}) and the SDSS DR3 data set and concluded that the Hubble law is linear up to small redshifts less than $0.3$ but nonlinear for higher redshift, which adds fuel to the cosmological debate.

\ifnum \value{separate}=1
\clearpage
\else
\fi

\citealt{Roy2000} explain the broadening of the spectral lines using DMS. They also found a critical source frequency below which no spectrum can be observed for a particular medium. The broadening due to multiple scattering is more than the shift due to a cosmological effect.

The generally accepted idea is that a galaxy is surrounded by background quasars. However, this idea runs counter to a statistical analysis of the background/foreground object probability in a small area distributed according to its position and average density in any line of sight. The main approach here is to demonstrate that two extragalactic objects with very different redshifts may in reality be physical neighbors.

Physical association of quasars and galaxies can also be explained using SMBH and AGN. An SMBH can grow in size up to billions of solar masses from a ``seed'', the growth process being dependent on the feeding mechanism from surrounding gas. Astronomers think that every galaxy has an extremely bright center region, referred to as an AGN, and it is thought that it is powered by an SMBH at its center. The most luminous of all of the AGNs are quasars. The recent discovery of J0313-1806 leads to a question about its rapid formation as it is thought to sit inside a galaxy with a very active region, with the host galaxy producing $200$ solar masses worth of stars per year. The SMBH that powers this quasar presumably formed just $670$ million years after the presumed expansion of spacetime began, begging the question of how a seed grew large enough to form a black hole, with ``direct collapse'' proposed as a possible origin.

One of the outstanding conundrums in astronomy today is how did SMBH, weighing millions to billions of times the mass of the Sun, get to be so huge so fast and so near the presumed beginning of the Universe. Recent observations of extragalactic objects do not appear to be consistent with the cosmological hypothesis that their redshifts arise from the expansion of spacetime. Observational evidence suggests that a galaxy in the early universe as mature as our galaxy is contradictory, a discrepancy that does not exist if high redshifts have a largely intrinsic component (\citealt{Fulton2012}).

\section{Summary and conclusions}
We have observed the existence of redshift periodicity in quasar-galaxy pair redshift data with transformation to a galaxy rest frame and in quasar and separately galaxy redshift data without transformation.  Our results directly contradict the proposition by other analysis that redshift periodicity does not exist, for example that any such observations are actually due to a selection effect of data binning (\citealt{Basu1978}), or due to noise in the fundamental periodicity within a particular radiation band. The most significant contribution of the present work is the analysis of the pair redshift data and its periodicity detection and estimation using an SVD based approach rather than the conventional periodogram FFT based approach. The applicability of the SVD approach, also used in \citealt{Mal2020} and \citealt{Mal2022} is further vindicated by the present results. SVD is therefore invaluable for astronomical data analysis. It is able to reveal hidden periodicities, leading to correct interpretation of the data.

\ifnum \value{separate}=1
\clearpage
\else
\fi

\citealt{Fulton2012} and \citealt{Fulton2018} have statistically demonstrated the existence of Karlsson peaks and quasar redshift periodicity using periodicity constraints to confine the hypothesised quasar ejection velocity while also significantly reducing false positive quasar family detections. This paper adds weight to their argument by independently confirming the existence of redshift quantisation in quasar, galaxy, and quasar-galaxy pair data sets. These new results also support the Hoyle-Narlikar model (\citealt{Hoyle1964, Hoyle1966}) and the Narlikar–Das cosmological model (\citealt{Narlikar1980}; \citealt{Burbidge2000}) in which quasars are ejected from galactic nuclei and in so doing define the true process of quasar and galaxy evolution.

Cosmologists are greatly interested in understanding the source of the Lyman-$\alpha$ forest and any related physical processes occurring in the environments surrounding galaxies. We posit that a key element of the ultimate elucidation of that physics is provided by VMH by virtue of its contention that a lower redshift quasar implies an older and larger mass whereas a higher redshift quasar implies a younger and smaller mass. In contrast to the standard model, the absorption features in the Lyman-$\alpha$ forest should exhibit peaks at multiples of the base periodic component (\citealt{Karlsson1971, Karlsson1973, Karlsson1990} and \citealt{Burbidge1968}). This paper confirms redshift periodicity of quasars, galaxies, and quasar-galaxy pairs, but 3D imaging may provide a more detailed understanding of quasars ejected from galactic nuclei. In our future work we will discuss the possible direct estimation of redshift periodicity within the DMS framework.


This paper demonstrates that, in contrast to the Standard Model (SM) of cosmology, redshift periodicity exists independently in each quasar, galaxy, or pair data set, and that periodicity exists in quasar and galaxy redshifts each independently. The recent observations just prior to and now pouring in from the James Webb Space Telescope (JWST) prompt us to make the following connections to the existence of extragalactic redshift periodicity.

\begin{enumerate}

\item\label{I-EverIncreasingRedshifts}{

\noindent
\textbf{Ever increasing redshifts:} The ever increasing maximum observed redshift is a huge problem for the SM, GN-z11 ($z > 10$; \cite{Jiang2021}) being just one of the more recent examples showing up prior to JWST and now with \cite {yan2022} reporting $z \approx 11-20$ in the first batch of JWST candidate objects. The problem does not exist for an alternative model that embraces redshift periodicity because, as per Sections 1 and 9 in \cite{Fulton2018} and elsewhere in the literature, a major portion of all high as well as mid-range redshifts is \textit{intrinsic}. Very high redshifts even for very distant objects are more than reasonable if the objects are simply examples of a quasar with very a large intrinsic redshift ejected from a very distant galaxy, the combined effects being the reason for the ever increasing measured redshifts.}

\item\label{I-HighRedshiftAndHighLuminoscity}{

\noindent
\textbf{High redshift and high luminosity:} The high redshift and extremely luminous quasars, such as GNz7q (\cite{Fujimoto2022}), are multiply problematic for SM, first for the reason stated in Item \ref{I-EverIncreasingRedshifts}, but also due to an outlandish luminosity, even to the extent of exceeding the otherwise generally accepted Eddington limit (\cite{Schindler2021}). As with the high redshift, the luminosity is problematic only if the object is evolving on its own at a distance corresponding to an SM cosmological redshift, as opposed to being at a much smaller distance and therefore also lower luminosity because it is evolving after ejection from a much closer but unaccounted parent galaxy.}

\item\label{I-MetallicityAndCarbonEvolution}{

\noindent
\textbf{Metallicity and carbon evolution:} The greater distances forced by insisting on cosmological redshifts interfere with the modeling and tracking of the evolution of high redshift quasars. \cite{Juarez2009} observed high redshift quasars with well-evolved metallicity but also an abundance of carbon relative to silicon and oxygen, neither of which are expected for young quasars.}

\item\label{I-TheLargestGalaxy}{

\noindent
\textbf{The largest galaxy:} \cite{Oei2022} have isolated what they have determined in projection to be the unambiguously largest giant radio galaxy (GRG) yet detected. A GRG is assumed to be hosted by another galaxy acting as its progenitor, and in this particular case both the GRG and the progenitor appear to be otherwise ordinary in terms of a long list of presumably relevant attributes, including an unambiguous and unspectacular host redshift, $z_{spec} = 0.24674$, and notably they reside in a low density environment. What is pointed out by \cite{Oei2022} and is of note in the current context is that neither very massive galaxies nor supermassive black holes are required to produce these giant entities, and this sits in conflict with rationalizations being made about presumably very large objects being generated in the very early Universe and of course also at very high redshift.}

\item\label{I-MilkyWayAndOtherFilaments}{

\noindent
\textbf{Milky Way and other filaments:} \cite{Yusef2022} compare the morphology and other attributes of filamentary structures in the center of the Milky Way with those found in radio galaxies in the intercluster medium of galaxy clusters. Due to the similarities between our local example and those far away and therefore much older galaxies, they argue that the underlying physical processes of the respective populations must be the same. If that is true, an obvious ramification is that extragalactic objects undergo the same evolutionary processes without regard to a particular location or epoch, as in an isotropic Universe. This in turn is consistent with an object having a redshift with an intrinsic component that decreases as the object evolves as part and parcel of a local phenomenon that is occurring everywhere and in every epoch.}

\item\label{I-DiversityAtHighRedshirtt}{

\noindent
\textbf{Diversity at high redshift:} \cite{Kartaltepe2022} have found a wide diversity in the structure and morphology of galaxies at $z = 3 - 9$ in early JWST observations. Their work includes repetitive visual classification, quantitative morphological determination, classification of type by percentage in different redshift ranges, and identification of trends in type and size that appear overall to exist over the full redshift range studied. The diversity and trend findings comport specifically with the contentions expressed in Item \ref{I-MilkyWayAndOtherFilaments} and generally with the thrust of all of these itemized conclusions.}

\item\label{I-RedshiftIsNotDistance}{

\noindent
\textbf{Redshift does not mean distance:} Redshift is not a distance indicator. In no case has an independent distance measurement been obtained for any object with $z \ge 1$, and this is also true for many extragalactic objects with $z < 1$. As stated in Item \ref{I-EverIncreasingRedshifts}, \cite{Fulton2018} have already determined that statistical samples (\cite{Schneider2010}; \cite{Shen2011}) of SDSS DR7 quasars do indeed have a large intrinsic redshift component, but at long last the literature from the wider community is publishing evidence that agrees with our contention that the long assumed redshift-distance relation has never been established. Based on the application of numerical methods to a joint X-ray and ultraviolet data set, \cite{Petrosian2022} have shown that a correlation between luminosity measurements in different wavelength regimes must be used with greater care to definitively demonstrate such a relationship. They state unequivocally that all of the correlation methods used to date suffer from a circularity argument, and we agree with that assessment.}

\end{enumerate}

\ifnum \value{separate}=1
\clearpage
\else
\fi

\normalem
\begin{acknowledgements}
The authors express their thanks to Prof. J. V. Narlikar of IUCCA for excellent guidance and suggestions provided during the research time. The authors want to express their special thanks to Martín L{\'o}pez-Corredoira for providing suggestions during discussions.  The authors also acknowledge Prof. Debasis Sengupta from ISI, Kolkata for his help in designing simulations and Ms. Susmita Nandi for her assistance in carrying out some simulations. First author wants to acknowledge Dr. Rajat Acharya and  Group Director Atual Shukla for their encouragement.
\end{acknowledgements}

\appendix                  

\section{SIMULATION OF DIFFERENT EFFECTS ON DATA SETS}
\label{sec:sim}

Here we simulate different effects on periodic data sets with different types of bias, such as clustering or the finite pixel size of a spectral graph. We assume that an ongoing process is repeating with a fixed length interval as mentioned below. The simulations performed are summarised in Table \ref{tab:table1} and described below.\\

\noindent
\textbf{Null hypotheses:} Generate a set of random  numbers  uniformly within a range of values, $0.036$ to $3.0$. Apply a periodicity detection method as mentioned in section $2.4$, and observe that the optimised bin width is $1.4811$ and no repeating pattern in histogram.

\noindent
\textbf{Aperiodic pattern:} Repeat the null hypothesis pattern with period lengths of $7, 22, 35, 60, 75, 80, 98,$ and $105$ in an aperiodic manner.
Analysis confirms non-existence of periodicity with a confidence level of $< 50\%$.

\noindent
\textbf{Periodic pattern with noise:} Repeat the null hypothesis pattern with a fixed length interval of $10$ and observe periodicity with a $95\%$ confidence level as indicated in Table \ref{tab:table1}.

\noindent
\textbf{Simple clustering effect:} A simple clustering effect is simulated by introducing another random process with no periodicity and with an amplitude higher than the periodic pattern with noise data set.
SVR1 is unable to detect a proper periodic length and instead observes a false periodicity due to the subharmonic amplitude being stronger than the fundamental peak.

\noindent
\textbf{Amplitude modulation clustering:} A clustering effect is simulated as a modulation in amplitude of a periodic data set.
Both SVR methods detect a fundamental period with a 95\% confidence level. The period is defined as the mean value of 1000 iterations of the fundamental period. SVR2 correctly detects periodicity in the simulated result, whereas SVR1 performs less well due to the noisy conditions.

\begin{table*}
\begin{center}

\caption{Summary of results of simulations of different effects on data sets.}

\label{tab:table1}

\begin{tabular}{lclrrr} \\
\hline\hline
Simulation                       & Bin                & Method & Known  & Confidence        & Confidence \\
                                 & Width              &        & Period & Interval          & Level      \\
\hline
Null hypothesis                  & $1.481$            & SVR1   & None     & No Dectection     &        $95\%$ \\
                                 &                    & SVR2   & None     & No Dectection     &        $95\%$\\
Aperiodic pattern                & $0.215$ to $0.367$ & SVR1   & None     & $[ 0.00 -  2.59]$ &   $< 50\%$ \\
                                 &                    & SVR2   & None     & $[ 0.00 -  2.59]$ &   $< 50\%$ \\
Periodic pattern with noise      & $0.206$ to $0.359$ & SVR1   & $10$   & $[ 4.96 - 10.09]$ &     $86\%$ \\
                                 &                    & SVR2   & $10$   & $[ 4.90 - 10.09]$ &     $95\%$ \\
Simple clustering effect         & $0.206$            & SVR1   & $10$   & $[ 3.09 -  6.18]$ &     $85\%$ \\
                                 &                    & SVR2   & $10$   & $[ 5.82 - 10.52]$ &     $95\%$ \\
Exponential variation clustering & $0.205$            & SVR1   & $10$   & $[10.00 - 15.09]$ &     $95\%$ \\
                                 &                    & SVR2   & $10$   & $[10.00 - 10.29]$ &     $95\%$ \\
Amplitude modulated clustering   & $0.051$ to $0.052$ & SVR1   & $10$   & $[ 1.66 - 10.02]$ &     $95\%$ \\
                                 &                    & SVR2   & $10$   & $[ 2.66 - 10.02]$ &     $95\%$ \\

\hline
\end{tabular}
\end{center}
\end{table*}

\ifnum \value{separate}=1
\clearpage
\else
\fi

\section{DYNAMIC MULTIPLE SCATTERING}
\label{sec:dms}

Below are the equations from \citealt{Roy2007} to distinguish between redshifts due to the DMS or Doppler mechanisms.

\begin{equation}
1 + z_{\text{DMS}} = \frac{\alpha' + (\alpha\alpha' - {\beta}^2){\delta}^2_{N}}{|\beta|}
\end{equation}
where
\begin{center}
${\delta}^2_{N} = (\frac{{\beta}^2}{{\alpha'}^2})^N {{\delta}_0^2} + \frac{\alpha'}{{\alpha'}^2 - {{\beta}^2}}$\\
\end{center}

\noindent
If ${\delta}_0$ is considered as due to Doppler broadening only, then

\begin{center}
${\delta}^2_{N} \simeq \frac{\alpha'}{{\alpha'}^2 - {\beta}^2}$
\end{center}

\noindent
Therefore,
\begin{equation}
1 + z_{\text{DMS}} = \frac{\alpha' + (\alpha\alpha' - {\beta}^2) [\frac{\alpha'}{{\alpha'}^2 - {\beta}^2}]}{|\beta|}
\end{equation}


\end{document}